\documentclass{article}
    \usepackage{microtype}
    \usepackage{graphicx}
    \usepackage{subfigure}
    \usepackage{booktabs} 
    \usepackage{amsmath}
    \usepackage{amssymb}
    \usepackage{amsfonts}
    \usepackage{float}
    \usepackage{xcolor}
    \usepackage{colortbl}
    \usepackage{multirow}
    \usepackage{mathtools}
    \usepackage{tabto}
    \usepackage{todonotes}
    \usepackage{hhline}
    \usepackage{enumitem}
    \usepackage{hyperref}

    \usepackage[accepted]{icml2019}
    \newtheorem{definition}{Definition}
\icmltitlerunning{RED-Attack: Resource Efficient Decision based Attack for Machine Learning}

\begin{document}

\twocolumn[
    \icmltitle{RED-Attack: Resource Efficient Decision-based Imperceptible \\ Attack for Machine Learning}
    
    
    \icmlsetsymbol{equal}{*}
    
    \begin{icmlauthorlist}
       \icmlauthor{Faiq Khalid}{equal,to}
     \icmlauthor{Hassan Ali}{equal,goo}
       \icmlauthor{Muhammad Abdullah Hanif}{to}
      \icmlauthor{Semeen Rehman}{to}
      \icmlauthor{Rehan Ahmed}{goo}
      \icmlauthor{Muhammad Shafique}{to}
   \end{icmlauthorlist}
    
   \icmlaffiliation{to}{Vienna University of Technology, Vienna, Austria}
 \icmlaffiliation{goo}{National University of Sciences and Technology, Islamabad, Pakistan}
    
    \icmlcorrespondingauthor{Faiq Khalid}{faiq.khalid@tuwien.ac.at}
    \vskip 0.2in
]


\printAffiliationsAndNotice{\icmlEqualContribution} 
\begin{abstract}
	Due to data dependency and model leakage properties, Deep Neural Networks (DNNs) exhibit several security vulnerabilities, e.g., training or inference data poisoning, backdoors, Trojans, model stealing, etc. Several security attacks, especially the black-box-based, exploited these vulnerabilities but most of them require output probability vector/distribution. These attacks can be mitigated by concealing the output probability vector/distribution. To address this limitation, decision-based attacks have been proposed which can estimate the model (gradient, classification behavior and other parameters) but they require several thousand queries to generate a single untargeted attack image. However, in real-time attacks, resources and attack time are very crucial parameters. Therefore, in resource constrained systems, e.g., autonomous vehicles where an untargeted attack can have a catastrophic effect, these attacks may not work efficiently. To address this limitation, we propose a resource efficient decision-based methodology which generates the imperceptible attack, i.e., the RED-Attack, for a given black-box model.
	
	The proposed methodology follows two main steps to generate the imperceptible attack, i.e., classification boundary estimation and adversarial noise optimization. Firstly, we propose a half-interval search-based algorithm for estimating a sample on the classification boundary using a target image and a randomly selected image from another class. Secondly, we propose an optimization algorithm which first, introduces small perturbation in some randomly selected pixels of the estimated sample. Then to ensure imperceptibility, it optimizes the distance between the perturbed and target samples. 
	
	For illustration, we evaluate it for CFAR-10 and German Traffic Sign Recognition (GTSR) using state-of-the-art networks. The experimental results show that it generates adversarial examples with better imperceptibility (96.1\%, 71.7\% and 203\% for perturbation norm, Structural Similarity Index and Correlation Coefficient, respectively, for limited iteration (1000)) as compared to Decision-based Attack.
\end{abstract}
%
\section{Introduction}\label{introduction}
Deep Neural Networks (DNNs) have revolutionized the computing paradigms but due to their inherent security vulnerabilities, i.e., training or inference data poisoning \cite{DBLP:journals/corr/abs-1811-01444}, backdoors \cite{biggio2018wild,gu2017badnets}, Trojans \cite{li2018hu,zou2018potrojan,chen2018detecting}, etc., they can lead to catastrophic effects in safety-critical applications, e.g., in autonomous driving \cite{stilgoe2018machine}. Several security attacks have been proposed that exploit these security vulnerabilities, however, the adversarial attacks have emerged as one of the most common and successful class of security attacks against DNNs and can be defined as the carefully-crafted imperceptible data corruptions to fool DNNs for misclassification \cite{goodfellow2014explaining}. The implementation and effectiveness of these attacks depend upon the underlying assumption for attacker's access to the DNNs, i.e., white-box and black-box scenarios.
\begin{figure*}[!t]
	\centering
	\includegraphics[width=1\linewidth]{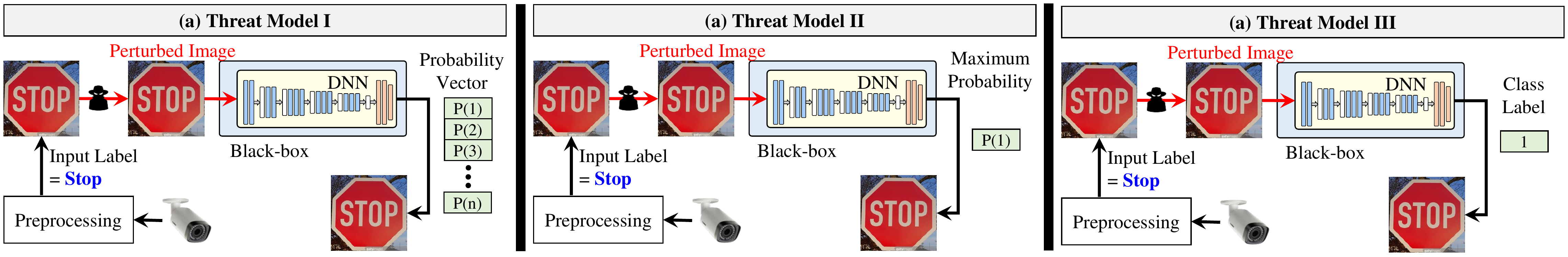}
	\caption{\textit{Threat Models: Different black-box attack and defense scenarios while assuming the black-box model and access to output classification probability vectors/distribution or labels}}
	\label{fig:Black-scenarios}
	\vskip -0.1in
\end{figure*}
\begin{figure*}[!t]
	\centering
	\includegraphics[width=1\linewidth]{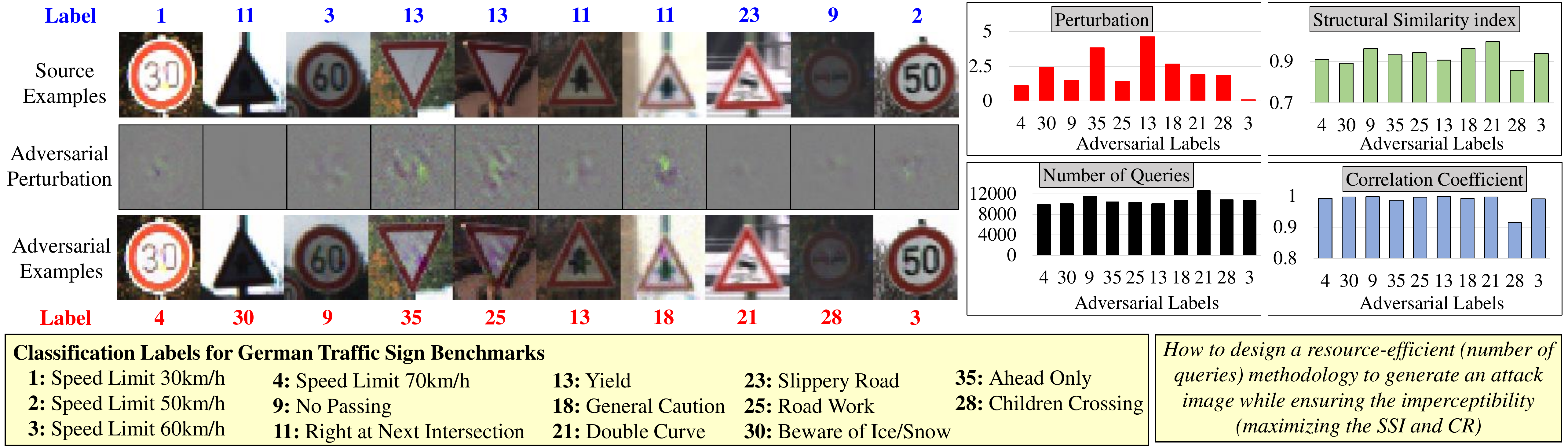}
	\caption{\textit{Adversarial Examples generated using Decision-based Un-targeted Attacks on German Traffic Sign Recognition benchmarks for the 10000 iterations. (Perturbation: strength of Adversarial noise, SSI: Structural Similarity Index; CR: Correlation Index )}}
	\label{fig:DF}
	\vskip -0.15in
\end{figure*}

\textbf{White-Box Attacks}: Most of the state-of-the-art attacks assume the white-box scenario in which the adversary has full knowledge of the DNN architecture and the corresponding parameters (weights and biases), e.g., Fast Gradient Sign Method (FGSM) \cite{goodfellow2014explaining}, iterative FGSM (I-FGSM) \cite{DBLP:journals/corr/KurakinGB16}, Jacobian Saliency Map Attack (JSMA)  \cite{DBLP:journals/corr/PapernotMJFCS15}, Carlini-Wagner (CW) Attack \cite{carlini2017towards}, One-Pixel Attack, Universal Adversarial Perturbation, DeepFool, PGD-Attack, etc. Although these attacks can generate imperceptible adversarial examples efficiently, \textit{the underlying assumption of access to the complete DNN model is impractical in most of the scenarios}.

\textbf{Black-Box Attacks}: Unlike white-box attacks, black-box attacks are not based on the aforementioned assumption and, therefore, limits the capability of the adversary because of the limited knowledge about the target system. Based on the available information in black-box scenarios, following can be the three possible threat models:
\begin{enumerate}
\itemsep0em 
    \item \textit{Threat Model I:} Adversary has access to the output probability vector/distribution while does not have access to the model parameters, as shown in Fig. \ref{fig:Black-scenarios}(a).
    \item \textit{Threat Model II:} Adversary has access only to the probability of the most probable class detected by the DNN classifier (in case of a classification problem), as shown in Fig. \ref{fig:Black-scenarios}(b).
    \item \textit{Threat Model III:} Adversary has access only to the final output of the system, i.e., the final class label in case of the a classification, as shown in Fig. \ref{fig:Black-scenarios}(c).
\end{enumerate}
Most of the state-of-the-art black-box attacks assume threat models I or II, e.g., Zeroth-Order Optimization-based Attack \cite{DBLP:conf/ccs/ChenZSYH17} and Gen-Attack \cite{DBLP:journals/corr/abs-1805-11090}. However, these attacks can be mitigated by \textit{concealing the information about the classification probability}, i.e., by considering Threat Model III (see Fig. \ref{fig:Black-scenarios}(c)). To address this limitation, decision-based attack \cite{DBLP:journals/corr/abs-1712-04248} has been proposed which utilizes the random search algorithm to estimate the classification boundary. Though decision-based attack can nullify the probability concealing-based defense, it possess the following limitations:
\begin{figure}[!t]
	\centering
	\includegraphics[width=1\linewidth]{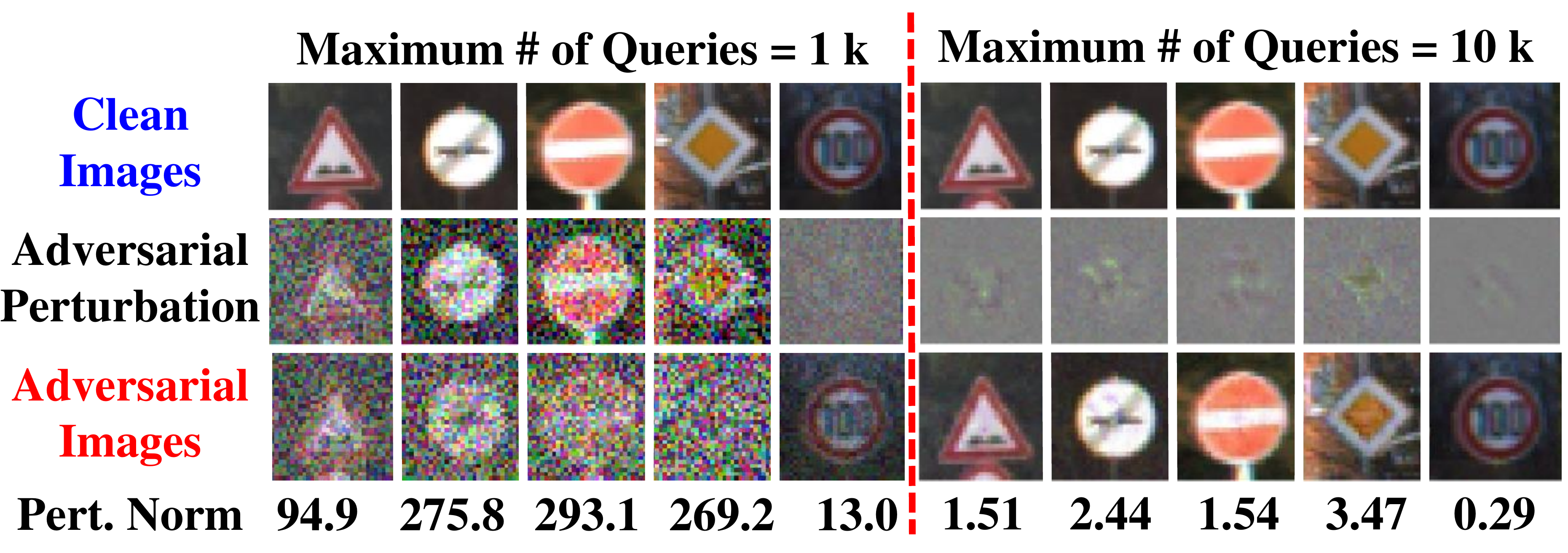}
	\caption{\textit{Adversarial Examples generated by the Decision-based attack with maximum iteration = 1000. This analysis shows that by reducing the number of iteration (limited resources), e.g., the perceptibility of the images increases, 1.51 to 94.9. Where $Pert. Norm = \sum{(Adversarial Image - Clean Image)^2}$}}
	\label{fig:MA2}
	\vskip -0.2in
\end{figure}
\begin{enumerate}
\itemsep0em 
    \item It requires multiple reference samples to generate a single adversarial example and thereby \textit{requires high memory/bandwidth resources}.
    \item The technique requires a large number of inferences/queries to generate a single adversarial example, for example, on average around 10000 for the examples illustrated in Fig. \ref{fig:DF}.
    \item In case of restricted number of allowed iterations (where each iteration can have multiple queries), the performance of the decision-based attack can reduces significantly, as illustrated in Fig. \ref{fig:MA2})
\end{enumerate}

\textbf{Associated Research Challenge:} Based on the above-mentioned limitations, we can safely conclude that decision-based attack cannot be applied to a resource and energy constraint systems, e.g., autonomous vehicles. Thus it raises a key research questions: \textit{How to reduce the number of queries for generating an adversarial example while maintaining the imperceptibility (maximizing the correlation coefficient and structural similarity index)?}
\subsection{Novel Contributions}
To address the above-mentioned research challenge, in this paper, we propose Resource-Efficient Decision-based methodology to generate an imperceptible adversarial attack under the threat model III. The underlying assumption for the proposed methodology is that it takes the pre-processed image and its corresponding class label from the black-box model. Then it iteratively generates the adversarial image multiple time, computes the corresponding classification label for each iteration and optimizes it by finding the closet adversarial image (on classification boundary) from input image, as shown in Fig. \ref{fig:Novel}. In a nutshell, this paper make the following contributions:
\begin{figure}[!t]
	\centering
	\includegraphics[width=1\linewidth]{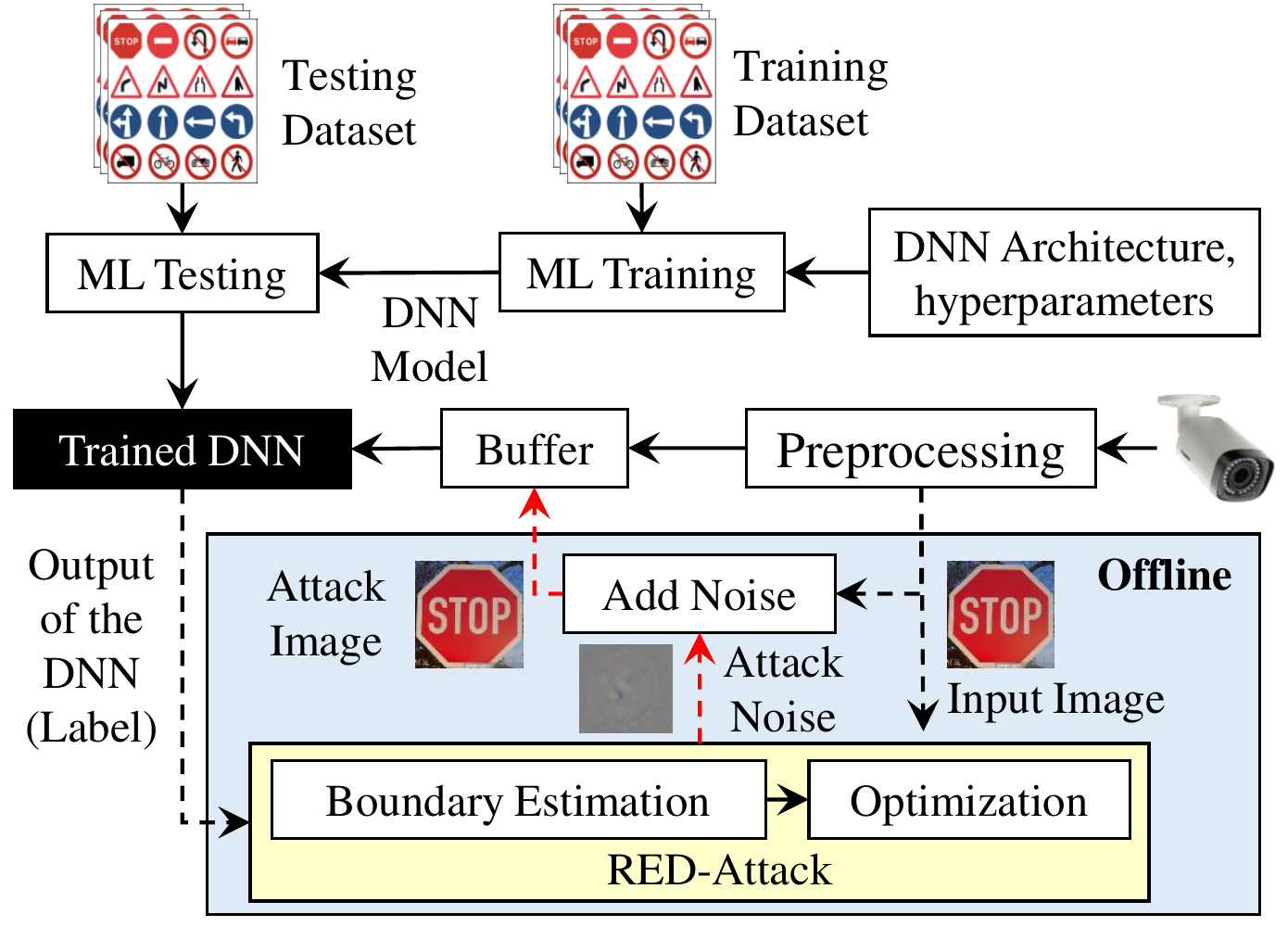}
	\caption{\textit{The proposed RED-Attack during typical design cycle for ML-based systems under the Threat Model III}}
	\label{fig:Novel}
	\vskip -0.15in
\end{figure}
\begin{enumerate}
\itemsep0em 
    \item To reduce the number of queries, we propose a single reference sample-based half-interval search algorithm to estimate the classification boundary. 
    \item To maximize the imperceptibility, we propose an optimization algorithm which combines the half-interval search algorithm with gradient estimation to identify the closest adversarial example (on boundary) from the input image. 
\end{enumerate}
To illustrate the effectiveness of the RED-Attack, we evaluate it for the CIFAR-10 and the GTSR dataset using state-of-the-art networks. The comparative analysis of results shows that the proposed approach successfully generates adversarial examples with much less perceptibility as compared to Decision-based Attack. We empirically show that, on average, the perturbation norm of adversarial images from their corresponding source images is decreased by 96.1 \%, while their SSI and CC with respect to the corresponding clean images is increased by 71.7 \% and 203 \%, respectively.

\section{Proposed Methodology} \label{MAD}
The proposed resource-efficient decision based attack methodology consists of two major steps: classification boundary estimation and optimization of attack image with respect to maximum allowed perturbations $\delta_{min}$.
\begin{figure}[!t]
	\centering
	\includegraphics[width=1\linewidth]{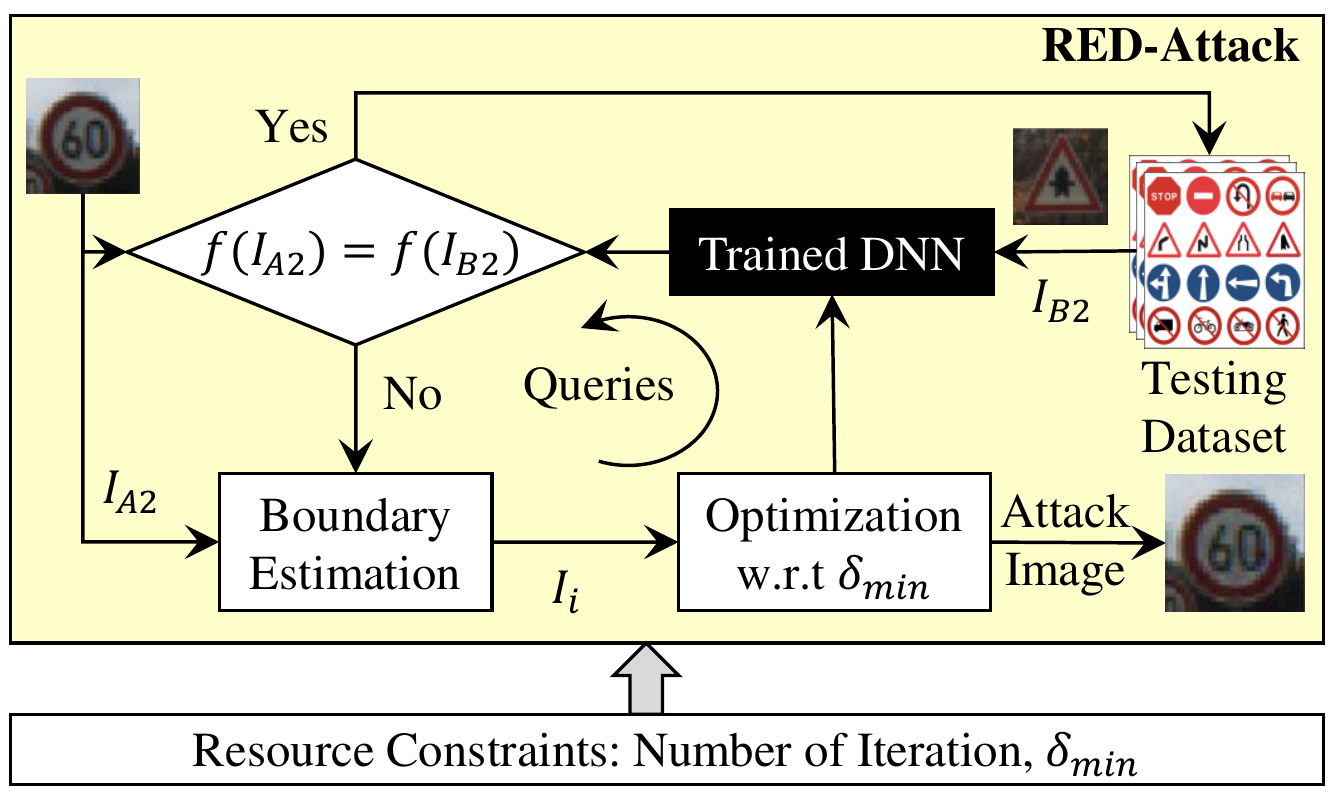}
	\caption{\textit{RED-Attack: Resource-Efficient Decision based methodology to generate on imperceptible attack while assuming the Threat Model III.}}
	\label{fig:RED-Attack}
	\vskip -0.15in
\end{figure}
\begin{figure*}[t!]
	\centering
	\includegraphics[width=1\linewidth]{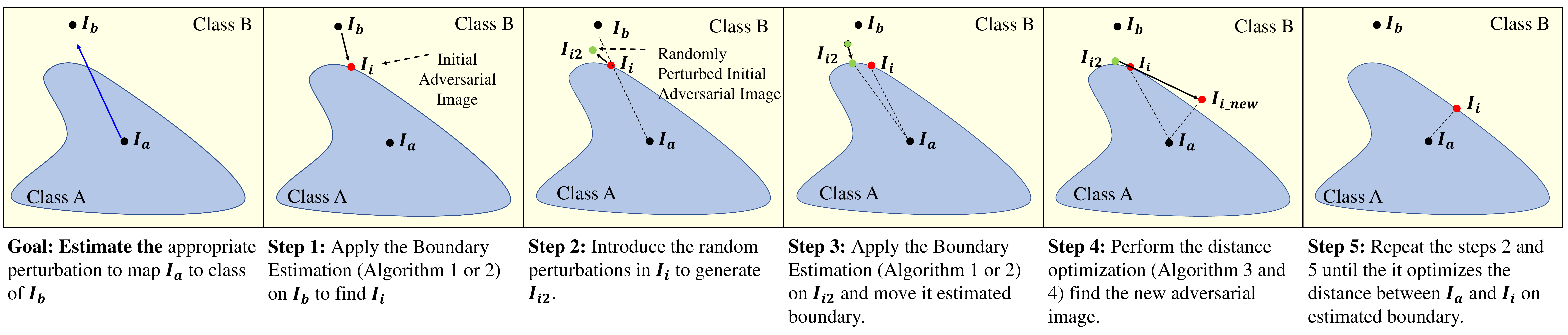}
	\caption{\textit{Proposed methodology to generate the optimized imperceptible hyper-sphere attacks}}
	\label{fig:proposed}
\end{figure*}
\begin{algorithm}[!t]
\caption{Boundary Estimation}
\label{algo:BE}
    \begin{algorithmic}
        \STATE \textbf{Inputs:} 
        \STATE \quad $I_{A2}$ = Source image; 
        \STATE \quad $I_{B2}$ = Sample Image from the target class;
        \STATE \quad $ \delta_{min}$ = Maximum Allowed Perturbation;
        \STATE \textbf{Output:}
        \STATE \quad $I_{i}$ = Adversarial Image; 
        \STATE Select a random sample Adversarial Image ($I_{i}$)
        \STATE Compute $k = f_{classifier}(I_i) $;
        \STATE Compute $\delta = max(I_{A2} - I_{i})$;
        \REPEAT 
        \IF { $f_{classifier}(I_A) \neq  k$ }
        \STATE Compute $I_{B2} = I_i$
        \ELSE
        \STATE Compute $I_{A2} = I_i$
        \ENDIF
        \STATE Compute $I_i = \frac{I_{A2} + I_{B2}}{2}$
        \STATE Compute $k = f_{classifier}(I_i) $;
        \STATE Compute $\delta = max(I_{A2} - I_{i})$;
        \UNTIL{$\delta > \delta_{min}$}
    \end{algorithmic}
\end{algorithm}
\subsection{Boundary Estimation}
The first step is to estimate the boundary for which we propose a half-interval search based algorithm that requires a target image and a reference image from any other class, as shown in Algorithm \ref{algo:BE} (See Step 1 of the Fig. \ref{fig:proposed}). 

\begin{definition}\label{def:BE}
Let $I_{A2}$, $I_{B2}$ and $ \delta_{min} $ be the source image (class: A), reference image (class: other than A) and maximum allowed estimation error. The goal of this algorithm is to find a sample which has tolerable distance (less than $\delta_{min}$) from the classification boundary and has label different from that of the source image, mathematically it can be defined as:

 $\exists\ I_{i}: f(I_{i})\ \neq\ f(I_{A2})\ \land\ max(I_{i}-I_{A2})\ \leq\ \delta_{min} $
 \vskip -0.15in
\end{definition}
To generate the appropriate $I_{i}$, the proposed algorithm perform the half-interval search using the source image ($I_{A2}$) and the reference image ($I_{B2}$). It first finds the half way point $I_{i}$ between $I_{A2}$ and $I_{B2}$ by computing the average of the two and then replaces $I_{A2}$ or $I_{B2}$ with $I_{i}$ depending upon the class in which $I_{i}$ falls. For examples, if the label of the half way point $I_{i}$ is class A then algorithm replaces $I_{A2}$ with $I_{i}$ and if its label is not A then the algorithm replaces $I_{B2}$ with $I_{i}$. The algorithm repeats this process until the maximum distance of $I_{i}$ from the $I_{A2}$ is less than $\delta_{min}$ and while ensuring the $f(I_{i})\ \neq f(I_{A2})$. The proposed boundary estimation can be used for targeted attack if we choose the reference image from the target class.
\begin{algorithm}[!t]
\caption{Gradient Estimation}
\label{algo:GE}
    \begin{algorithmic}
        \STATE \textbf{Inputs:} 
        \STATE \quad $I_{A}$ = Source image; 
        \STATE \quad $I_{B}$ = Sample Image from the target class;
        \STATE \quad $I_i$ = Adversarial Image;
        \STATE \quad $n$ = Number of pixels to perturb;
        \STATE \quad $\theta$ = Relative Perturbation in each pixel;
        \STATE \textbf{Output:}
        \STATE \quad $I_{i2}$ = Noisy Adversarial Image;
        \STATE \quad $g$ = Sign of the Gradient;
        \STATE Select a random sample Adversarial Image ($I_{i}$);
        \STATE Compute $I_0 = zeros$;
        \STATE Set randomly selected \(n\) pixels of $I_0$ to the maximum value a pixel can have;
        \STATE Compute $I_{i2} = I_i + \theta \times I_0$; and Compute $I_{i2}$
        \STATE Compute $d2 = \sum (I_{i2} - I_A)^2$; and $d1 = \sum (I_{i} - I_A)^2$;
        \IF { $d2 > d1$ }
        \STATE Compute $g = -1$;
        \ELSIF { $d2 < d1$ }
        \STATE Compute $g = 1$;
        \ELSE
        \STATE Compute $g = 0$;
        \ENDIF
    \end{algorithmic}
\end{algorithm}

\subsection{Optimization of Attack Image}
The second step of the proposed methodology is to optimize noise on the sample $I_{i}$ (output of the boundary estimation). We propose to incorporate the adaptive update in the zeroth-order stochastic algorithm to efficiently optimize it. 
\begin{algorithm}[!t]
\caption{Efficient Update}
\label{algo:EU}
    \begin{algorithmic}
        \STATE \textbf{Inputs:} 
        \STATE \quad $I_{A2}$ = Copy of the Source image; 
        \STATE \quad $I_{B}$ = Sample Image from the target class;
        \STATE \quad $I_i$ = Adversarial Image;
        \STATE \quad $I_{i2}$ = Randomly Perturbed Adversarial Image;
        \STATE \quad $g$ = Gradient Sign;
        \STATE \quad $j$ = Maximum Jump;
        \STATE \textbf{Output:}
        \STATE \quad $I_{i,new}$ = A new instance of adversarial example;
        \STATE Compute $\Delta = g \times (I_{i2} - I_i)$; and $\lambda = j$;
        \STATE Compute $I_{i,new} = I_i + \lambda \times \Delta$;
        \REPEAT
        \STATE Compute $\lambda = \frac{\lambda}{2}$; and $I_{i,new} = I_i + \lambda \times \Delta$;
        \UNTIL{$\sum (I_{i,new} - I_{A})^2 > \sum (I_{i} - I_{A})^2$}
        \IF{$\sum (I_{i,new} - I_{A})^2 > \sum (I_{i} - I_{A})^2$}
        \STATE Compute $I_{i,new} = I_i$;
        \ENDIF
    \end{algorithmic}
\end{algorithm}
\begin{figure*}[t!]
	\centering
	\includegraphics[width=1\linewidth]{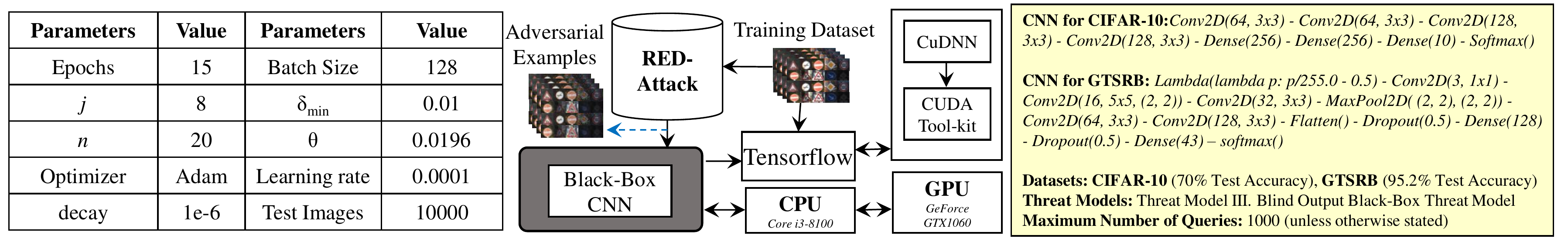}
	\caption{\textit{Experimental setup and tool flow for evaluating the proposed RED-Attack}}
	\label{fig:Tool}
\end{figure*}
\begin{definition}\label{def:opt}
Let $I_{A2}$, $I_{B2}$, $ \delta_{min} $ and $I_{i}$ be the source image (class: A), reference image (class: other than A), maximum allowed estimation error and perturbed image, respectively. The goal of this algorithm is to minimize distance of $I_{i}$ from $I_{A2}$ while ensuring that it has label different than the source image, mathematically it can be defined as:

 $\forall\ I_{i}\ min(I_{i}-I_{A2}): f(I_{i})\ \neq f(I_{A2})\ $
\end{definition}
To achieve this goal, we first identify the sign of the gradient for estimating the boundary behavior (minima). For example, if the sign is positive then we continue moving in the same direction; otherwise we switch the direction, as shown in Algorithm \ref{algo:GE}. Once the direction is identified, the next key challenge is to select the appropriate hop size. Therefore, to address this challenge, we propose an algorithm (Algorithm \ref{algo:EU}) which introduces adaptive hop size by applying the half-interval update. For example, for updating, it starts with maximum hop size and then it reduces it by half. The algorithm repeats itself until it finds the local minima, illustrated by Step 3-5 of Fig. \ref{fig:ED}.


\section{Experimental Results and Discussions} \label{MAD}

\subsection{Experimental Setup}
To demonstrate the effectiveness of the proposed RED-Attack, we evaluated several un-targetd attacks on CIFAR10 and GTSR for state-of-the-art networks (Fig. \ref{fig:Tool}).  
\subsection{Experimental Analysis}
\begin{figure}[!t]
	\centering
	\includegraphics[width=1\linewidth]{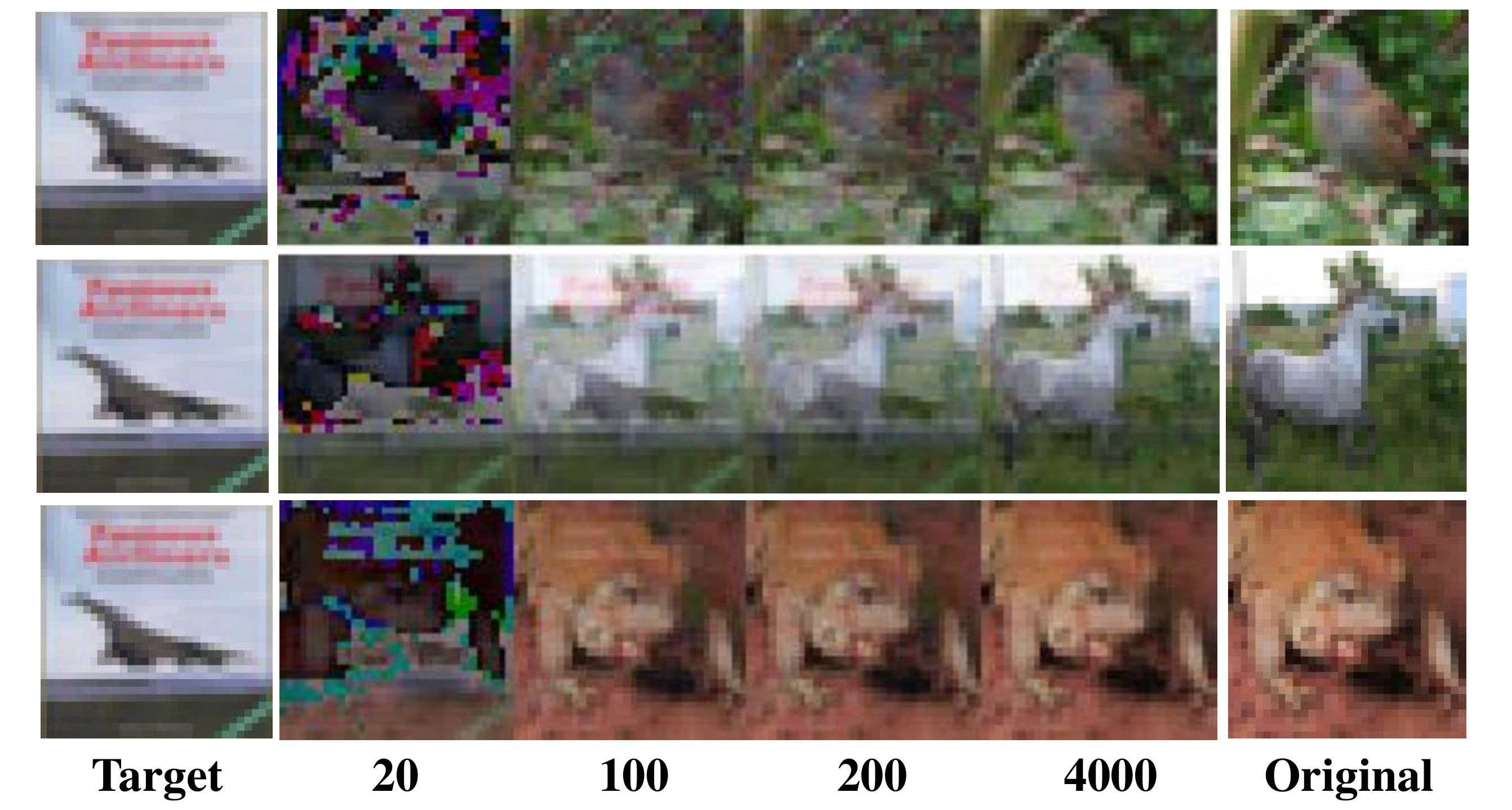}
	\caption{\textit{Visualizing the adversarial examples at various query counts.}}
	\label{fig:VT}
	\vskip -0.1in
\end{figure}

The imperceptibility of the adversarial image improves iteratively  by minimizing its distance from the source image, as shown in Fig. \ref{fig:VT}. Form the analysis of this figure, we identify the following key observation: The adversarial image generated in first few iterations are not imperceptible but not even recognizable but over time with help of query and optimization algorithm it achieves the imperceptibility. Thus, in black-box attack, if we limit the number of queries then its imperceptibility decreases drastically. 
\subsection{Effects of Hyper-parameters on perceptibility (d)}
We measure the distance of adversarial image from its source source image using the L2-Norm of perturbation which is defined as the pixel-wise sum of \( (Source Image - Adversarial Image)^2 \). We monitor the effect of changing three different hyper-parameters, introduced in our attacks, on the distance of the adversarial examples from their corresponding source examples.
\subsubsection{Experimental Setup}
\begin{enumerate}[leftmargin=*]
\itemsep0em 
    \item \(\delta_{min}\) - measures the maximum tolerable error, in terms of each pixel value, while computing the boundary point in Algorithm 1. Unless, otherwise stated, the typical value we use for \(\delta_{min}\) is 1.
    \item \(n\) - defines the number of pixels, randomly selected to be perturbed in each iteration in order to estimate the gradient of the distance of the adversarial example from the source example (Algorithm 2). Unless otherwise stated, the typical value we use for \(n\) is 5.
    \item \(\theta\) - defines the magnitude of noise added in each of the \(n\) randomly selected pixels relative to the maximum value, a pixel can have. Unless otherwise stated, the typical value we use for \(\theta\) is 5.
\end{enumerate}

\begin{figure*}[t!]
	\centering
	\includegraphics[width=1\linewidth]{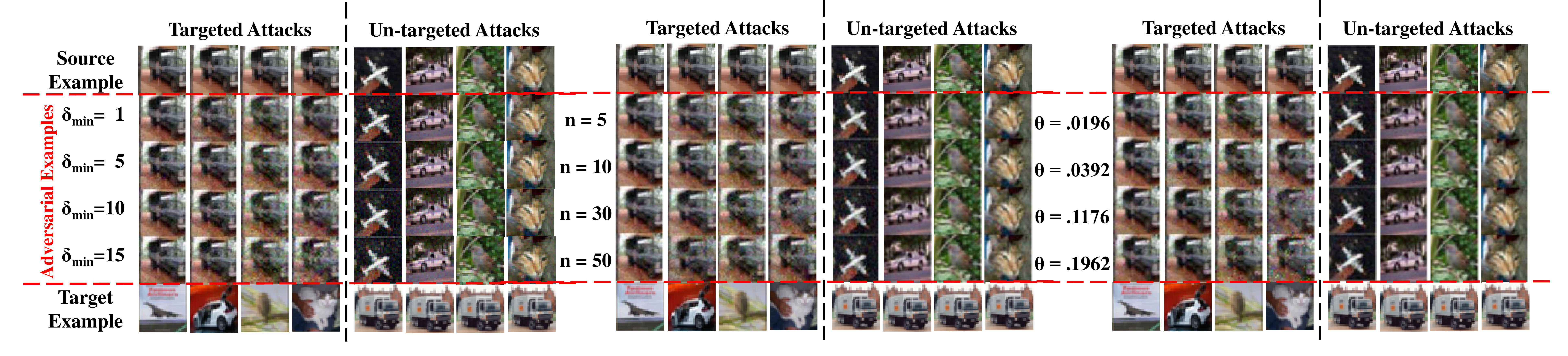}
	\caption{\textit{Adversarial examples generated by performing both targeted and un-targeted attacks for various values of \(\delta_{min}\), \(n and \theta\). The corresponding source/victim examples and the examples of the target class are given for each case. All the adversarial examples generated were successfully classified to be of the target class for targeted attacks and were mis-classified for the un-targeted attacks.}}
	\label{fig:CA}
	\vskip -0.1in
\end{figure*}
\begin{figure*}[t!]
	\centering
	\includegraphics[width=1\linewidth]{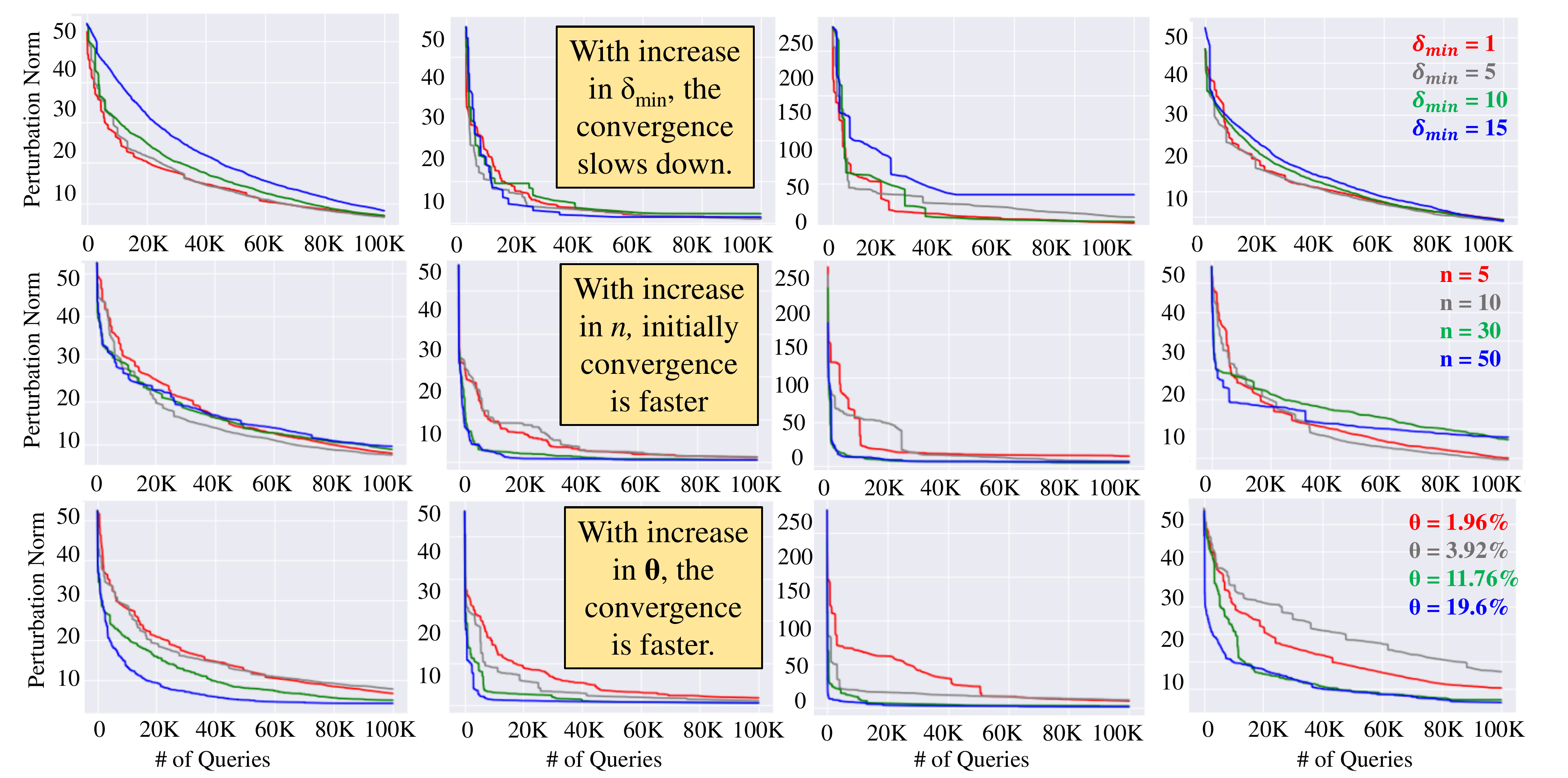}
	\caption{\textit{The trends of Perturbation norm (d) from the corresponding source examples over time for various values of \(\delta_{min}\), \(n  and \theta = 5\). (From left to right) Mis-classification attack on Airplane, Car, Bird and Cat. The initial target image was chosen to an image of a truck.}}
	\label{fig:CU}
\end{figure*}
   \begin{figure}[t!]
	\centering
	\includegraphics[width=1\linewidth]{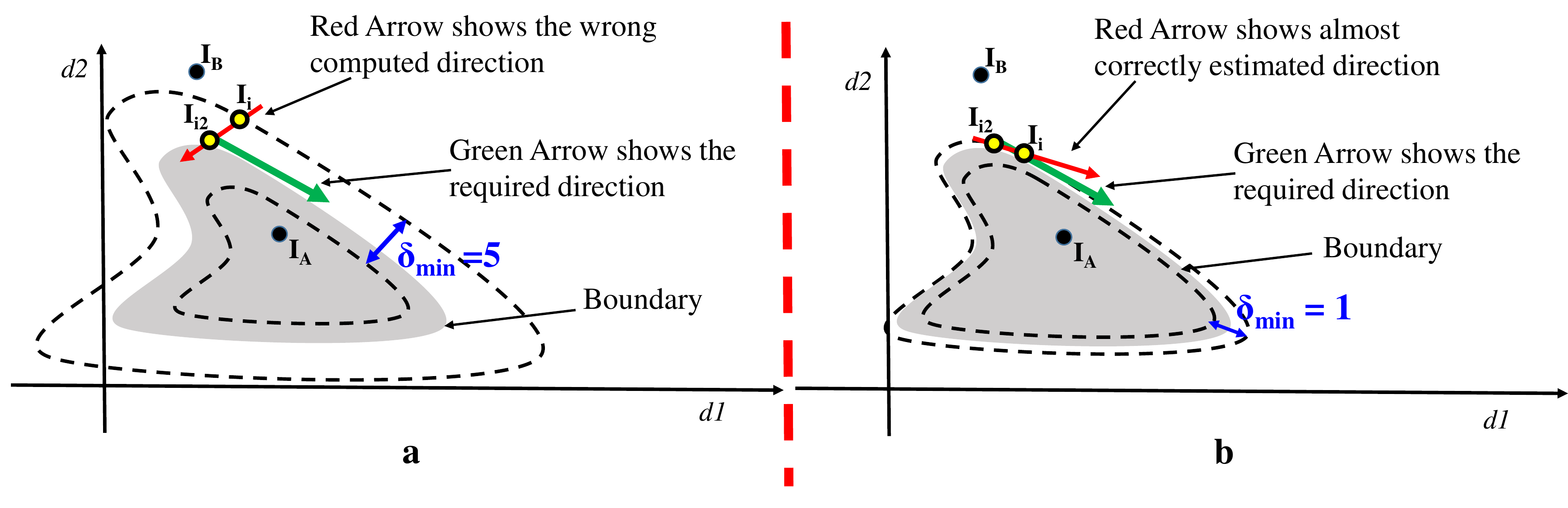}
	\caption{\textit{Illustrating the effects of changing \(\delta_{min}\) on the gradients. a) Incorrect estimation  for large \(\delta_{min}\). b) Close approximation for small  \(\delta_{min}\)}}
	\label{fig:ED}
	\vskip -0.1in
\end{figure}
\subsubsection{Experimental Analysis}
Figures \ref{fig:CA} and \ref{fig:CU} shows the source examples that we use for targeted and un-targeted attacks along with the initial target examples and the generated adversarial examples.
 
\begin{enumerate}[leftmargin=*]
\itemsep0em 
    \item As \(\delta_{min}\) increases, the quality of the adversarial example at a given query count decreases, due to the increase in its distance from the source example.
    This is because the larger value of \(\delta_{min}\) results in an imprecise boundary point, which in turn may result in an incorrect value of the estimated gradient, as shown in Figure \ref{fig:ED}. However, smaller value of \(\delta_{min}\) results in a gradient direction, closer to the required direction (see Figure \ref{fig:ED}b).
\begin{figure*}[t!]
	\centering
	\includegraphics[width=1\linewidth]{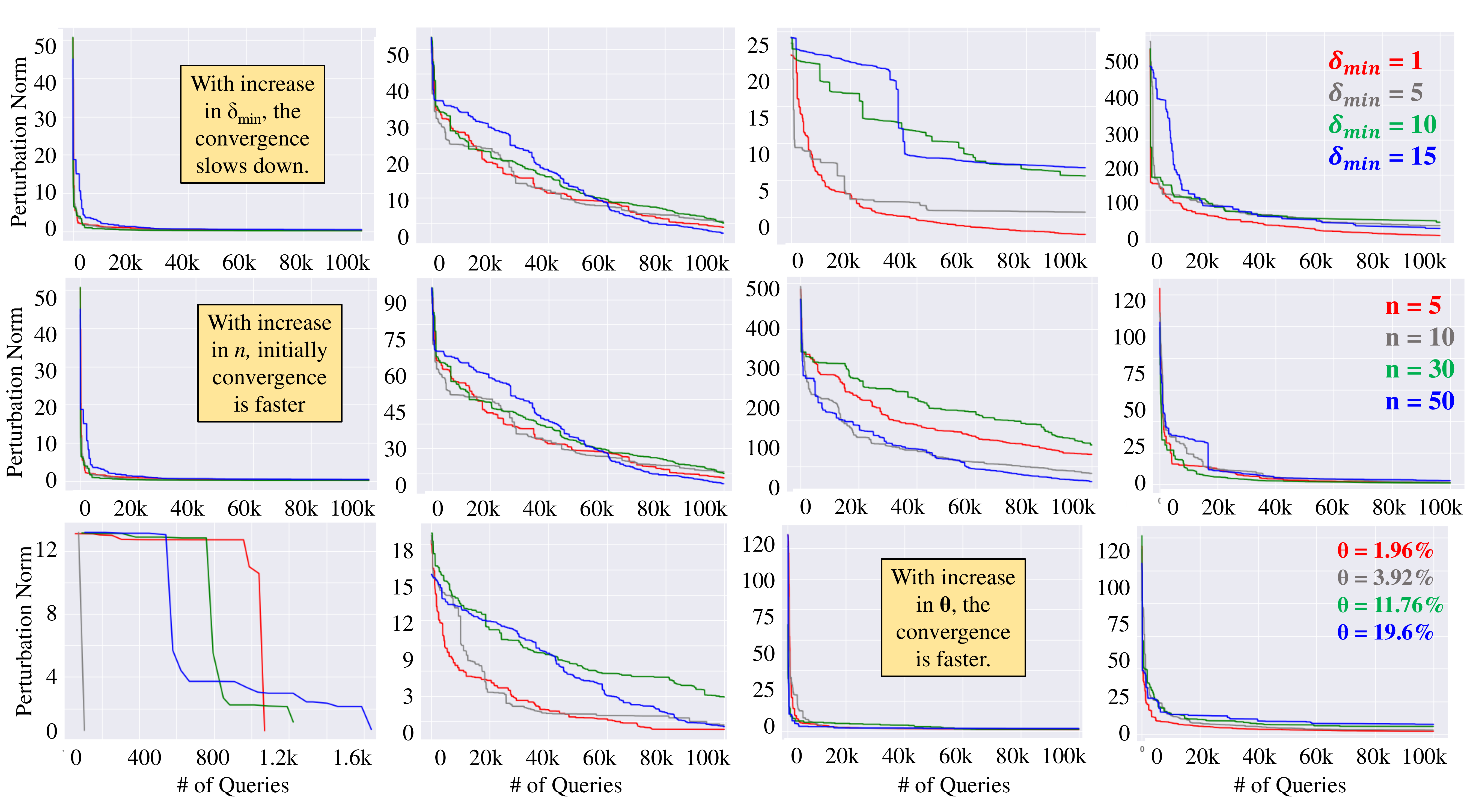}
	\caption{\textit{The trends of distance (d) of the adversarial example from the corresponding source examples as the algorithm progresses for various values of \(\delta_{min}\), \(n and \theta\) on German Traffic Sign Dataset. The source image, target image and the generated adversarial image are shown above each plot.}}
	\label{fig:GU}
	\vskip -0.1in
\end{figure*}
    
    \item A larger value of \(n\), initially results in a faster convergence. The reason is that we only need to estimate an overall trend of the boundary at initial stages. Estimating the update direction for the adversarial example by perturbing a large number of values at once helps achieve better results. However, Large \(n\) is highly vulnerable to divergence as the attack progresses. This observation suggests that the attack can be significantly improved by changing the number of pixels perturbed, as the algorithm progresses in an adaptive manner.
    
    \item Similar trend is observed  with the changes in \(\theta\) because estimating the gradients by introducing large perturbations in the image give an overall trend of the boundary instead of more precise localized gradients, as given by small perturbations. This in turn helps the algorithm to initially converge faster. However, small values of \(\theta\) give a more stable convergence towards the solution.
    
\end{enumerate}
\begin{figure*}[t!]
	\centering
	\includegraphics[width=1\linewidth]{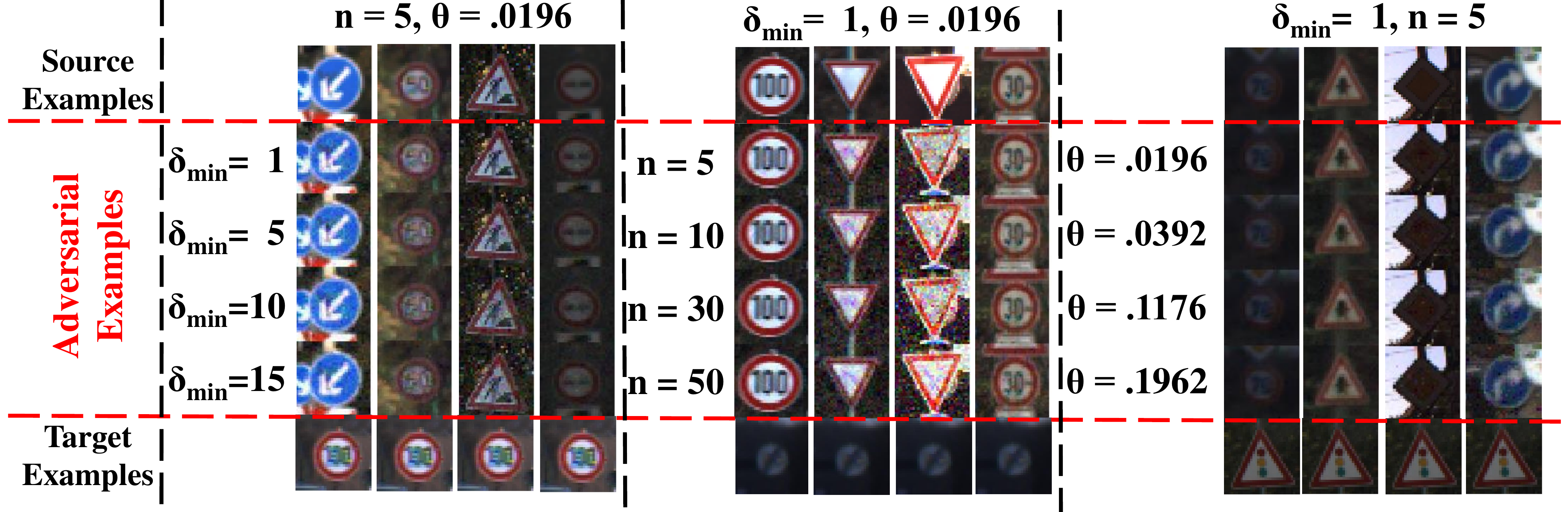}
	\caption{\textit{Adversarial examples generated by performing un-targeted attacks on the model trained for German Traffic Sign Classification for various values of \(\theta\), \(n\) and  \(\delta_{min}\). The corresponding source/victim examples along with the images of another class (target examples) specifying initial direction are given for each case. All the adversarial examples generated were successfully miss-classified by the classifier. Unless otherwise stated, \(\delta_{min}\) = 0.01, \(n\)=20, \(\theta\) = 0.0196. The maximum number of queries used to generate the images is \(10^5\)}}
	\label{fig:FU}
	\vskip -0.1in
\end{figure*}
\begin{figure*}[t!]
	\centering
	\includegraphics[width=1\linewidth]{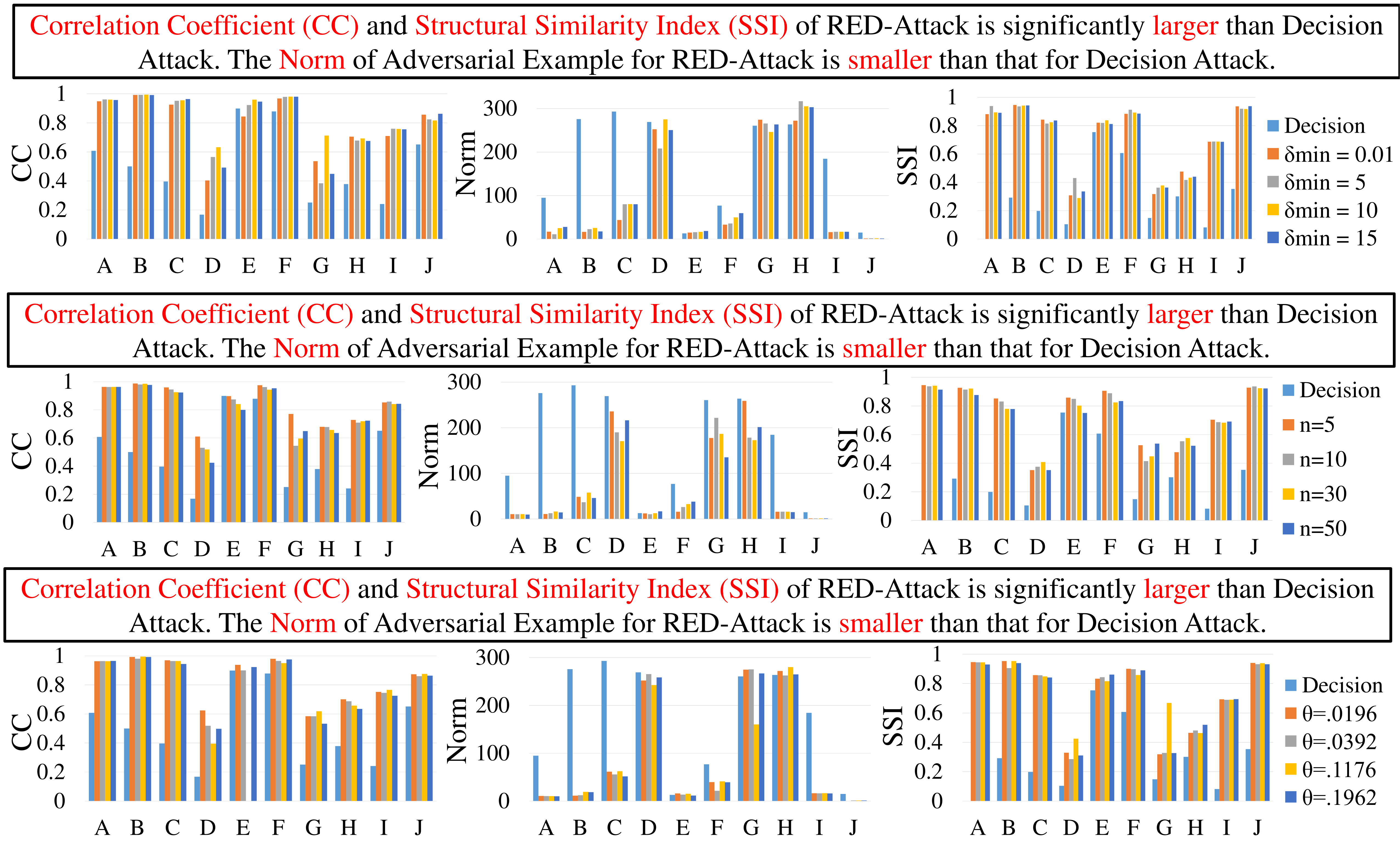}
	\caption{\textit{Comparison of RED-Attack for different images (see) with Decision-based Attack for various values of \(\delta_{min}\), \(n\) and \(\theta\). The maximum number of allowed queries \(q_{max} = 1000\). The metric used are Correlation coefficient (CC), Perturbation Norm (Norm) and Structural Similarity Index (SSI) computed against Adversarial images and the corresponding source images.}}
	\label{fig:comparison_all}
	\vskip -0.2in
\end{figure*}
\begin{figure*}[t!]
	\centering
	\includegraphics[width=1\linewidth]{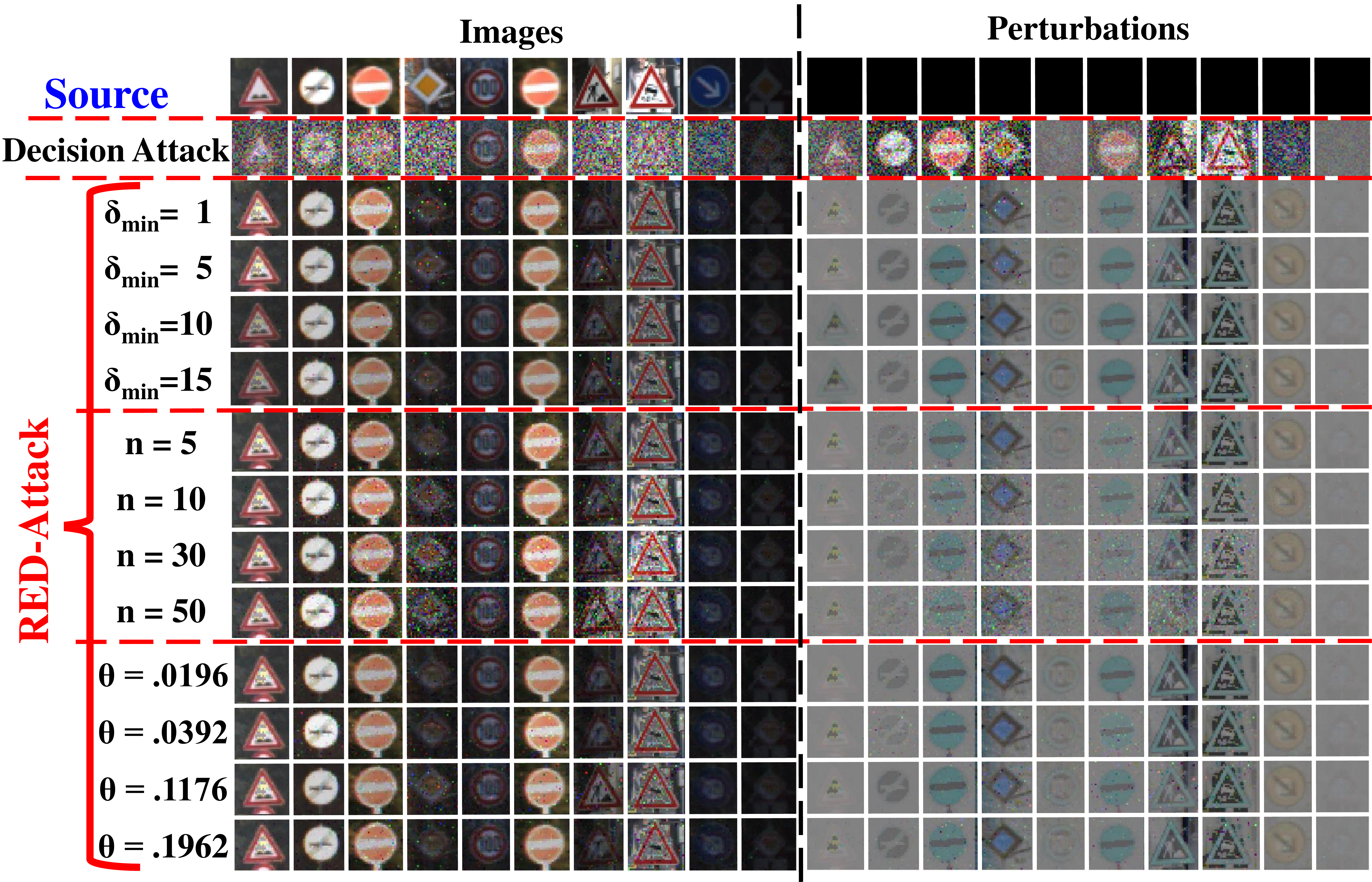}
	\caption{\textit{Comparison of the adversarial images generated by an untargeted Decision-based attack and those generated by an untargeted RED-Attack for various values of \(\delta_{min}\), \(n\) and \(\theta\). Please note that \(\delta_{min}\) = 0.01, \(n\) = 20 and \(\theta\) = 0.0196, unless otherwise stated. The maximum number of queries is 1000 for each adversarial image. }}
	\vskip -0.1in
	\label{fig:FC}
\end{figure*}
\subsection{German Traffic Sign Data set}
In order to find out the generality of the trends shown in Figure \ref{fig:CU}, we perform a similar analysis on GTSR. In this experiment, we randomly select any image from the test data and generate adversarial example for fixed number of queries. We compute perturbation norm (d) from their corresponding source images and compare different approaches based on these findings.

The results are shown in Figure \ref{fig:GU}. The complete set of adversarial examples along with the corresponding source images is shown in Figure \ref{fig:FU}. 

\subsubsection{Key Observations and Insights}
\begin{enumerate}[leftmargin=*]
\itemsep0em 
    \item Generally, the effect of changing hyper-parameters on the perturbation norm of the adversarial example is almost similar for GTSRB dataset and CIFAR-10.
    
    \item We observe that the adversarial examples for untargeted attack against GTSRB classifier converge much faster as compared to the CIFAR-10 dataset. We attribute this to much larger number of classes in GTSRB dataset as compared to CIFAR-10 dataset.
    
    \item As was observed in case of CIFAR-10, the attack can significantly be improved by adaptively changing the hyper-parameters as the attack progresses.
    
\end{enumerate}

\section{Comparison with the state-of-the-art} \label{MAD}

We compare our results with the state-of-the art Decision-based Attack. We use the implementation provided in an open source benchmark library, FoolBox \cite{DBLP:journals/corr/RauberBB17}. We limit the maximum number of queries to 1000 and evaluate our attack for different values of \(\delta_{min}\), \(n\) and \(\theta\). To compare our results with the Decision-based Attack, we use three different metrics i.e. correlation coefficient (CC), the squared L2-Norm (Pert. Norm) and the Structural Similarity Index (SSI) of the adversarial image with respect to the source image. Our results are shown in Figure \ref{fig:comparison_all}.

It can be seen from the figure that the adversarial examples produced by RED-Attack in this settings are significantly superior to those produced by the Decision Attack. The reason is binary stepping while searching for the boundary point and efficient update process while computing a new instance of adversarial example.

Here, we want to emphasize that in the long run, the decision-based attack eventually surpasses the RED-Attack e.g. if the query efficiency is not much of a concern or the number of maximum queries is limited to \(10^5\) instead of \(10^3\), the adversarial examples found by the decision-based attacks are better than those found by the RED-Attack.
\section{Conclusion}\label{conclusion}
In this paper, we proposed a novel Resource-Efficient Decision-based Imperceptible Attack (RED-Attack). It utilizes a half-interval search-based algorithm to estimate the classification boundary and an efficient update mechanism to boost the convergence of an adversarial example for decision-based attacks, in query limited settings. To illustrate the effectiveness of the RED-Attack, we evaluated it for the CIFAR-10 and the GTSRB dataset using multiple state-of-the-art networks. We limited the maximum number of queries to 1000 and showed that the state-of-the-art decision-based attack is unable to find an imperceptible adversarial example, while the RED-Attack arrives at a sufficiently imperceptible adversarial example within the predefined number of queries. Further, we empirically showed that, on average, the perturbation norm of adversarial images (from their corresponding source images) decreased by 96.1 \%, while their Structural Similarity Index and Correlation (with respect to the corresponding clean images) increased by 71.7 \% and 203 \%, respectively.
\bibliographystyle{icml2019}
\bibliography{bib/bibliography}

\begin{thebibliography}{15}
\providecommand{\natexlab}[1]{#1}
\providecommand{\url}[1]{\texttt{#1}}
\expandafter\ifx\csname urlstyle\endcsname\relax
  \providecommand{\doi}[1]{doi: #1}\else
  \providecommand{\doi}{doi: \begingroup \urlstyle{rm}\Url}\fi

\bibitem[Alzantot et~al.(2018)Alzantot, Sharma, Chakraborty, and
  Srivastava]{DBLP:journals/corr/abs-1805-11090}
Alzantot, M., Sharma, Y., Chakraborty, S., and Srivastava, M.~B.
\newblock Genattack: Practical black-box attacks with gradient-free
  optimization.
\newblock \emph{CoRR}, abs/1805.11090, 2018.
\newblock URL \url{http://arxiv.org/abs/1805.11090}.

\bibitem[Biggio \& Roli(2018)Biggio and Roli]{biggio2018wild}
Biggio, B. and Roli, F.
\newblock Wild patterns: Ten years after the rise of adversarial machine
  learning.
\newblock \emph{Pattern Recognition}, 84:\penalty0 317--331, 2018.

\bibitem[Brendel et~al.(2017)Brendel, Rauber, and
  Bethge]{DBLP:journals/corr/abs-1712-04248}
Brendel, W., Rauber, J., and Bethge, M.
\newblock Decision-based adversarial attacks: Reliable attacks against
  black-box machine learning models.
\newblock \emph{CoRR}, abs/1712.04248, 2017.
\newblock URL \url{http://arxiv.org/abs/1712.04248}.

\bibitem[Carlini \& Wagner(2017)Carlini and Wagner]{carlini2017towards}
Carlini, N. and Wagner, D.
\newblock Towards evaluating the robustness of neural networks.
\newblock In \emph{2017 IEEE Symposium on Security and Privacy (SP)}, pp.\
  39--57. IEEE, 2017.

\bibitem[Chen et~al.(2018)Chen, Carvalho, Baracaldo, Ludwig, Edwards, Lee,
  Molloy, and Srivastava]{chen2018detecting}
Chen, B., Carvalho, W., Baracaldo, N., Ludwig, H., Edwards, B., Lee, T.,
  Molloy, I., and Srivastava, B.
\newblock Detecting backdoor attacks on deep neural networks by activation
  clustering.
\newblock \emph{arXiv preprint arXiv:1811.03728}, 2018.

\bibitem[Chen et~al.(2017)Chen, Zhang, Sharma, Yi, and
  Hsieh]{DBLP:conf/ccs/ChenZSYH17}
Chen, P., Zhang, H., Sharma, Y., Yi, J., and Hsieh, C.
\newblock {ZOO:} zeroth order optimization based black-box attacks to deep
  neural networks without training substitute models.
\newblock In \emph{Proceedings of the 10th {ACM} Workshop on Artificial
  Intelligence and Security, AISec@CCS 2017, Dallas, TX, USA, November 3,
  2017}, pp.\  15--26, 2017.
\newblock \doi{10.1145/3128572.3140448}.
\newblock URL \url{https://doi.org/10.1145/3128572.3140448}.

\bibitem[Goodfellow~et al.(2014)]{goodfellow2014explaining}
Goodfellow~et al., I.~J.
\newblock Explaining and harnessing adversarial examples.
\newblock \emph{arXiv preprint arXiv:1412.6572}, 2014.

\bibitem[Gu et~al.(2017)Gu, Dolan-Gavitt, and Garg]{gu2017badnets}
Gu, T., Dolan-Gavitt, B., and Garg, S.
\newblock Badnets: Identifying vulnerabilities in the machine learning model
  supply chain.
\newblock \emph{arXiv preprint arXiv:1708.06733}, 2017.

\bibitem[Khalid~et al.(2018)]{DBLP:journals/corr/abs-1811-01444}
Khalid~et al., F.
\newblock Fademl: Understanding the impact of pre-processing noise filtering on
  adversarial machine learning.
\newblock \emph{CoRR}, 2018.
\newblock URL \url{http://arxiv.org/abs/1811.01444}.

\bibitem[Kurakin~et al.(2016)]{DBLP:journals/corr/KurakinGB16}
Kurakin~et al., A.
\newblock Adversarial examples in the physical world.
\newblock \emph{CoRR}, abs/1607.02533, 2016.

\bibitem[Li et~al.(2018)Li, Yu, Ning, Wang, Wei, Wang, and Yang]{li2018hu}
Li, W., Yu, J., Ning, X., Wang, P., Wei, Q., Wang, Y., and Yang, H.
\newblock Hu-fu: Hardware and software collaborative attack framework against
  neural networks.
\newblock \emph{arXiv preprint arXiv:1805.05098}, 2018.

\bibitem[Papernot~et al.(2015)]{DBLP:journals/corr/PapernotMJFCS15}
Papernot~et al., N.
\newblock The limitations of deep learning in adversarial settings.
\newblock \emph{CoRR}, abs/1511.07528, 2015.

\bibitem[Rauber et~al.(2017)Rauber, Brendel, and
  Bethge]{DBLP:journals/corr/RauberBB17}
Rauber, J., Brendel, W., and Bethge, M.
\newblock Foolbox v0.8.0: {A} python toolbox to benchmark the robustness of
  machine learning models.
\newblock \emph{CoRR}, 2017.
\newblock URL \url{http://arxiv.org/abs/1707.04131}.

\bibitem[Stilgoe(2018)]{stilgoe2018machine}
Stilgoe, J.
\newblock Machine learning, social learning and the governance of self-driving
  cars.
\newblock \emph{Social studies of science}, 48\penalty0 (1):\penalty0 25--56,
  2018.

\bibitem[Zou et~al.(2018)Zou, Shi, Wang, Li, Song, and Wang]{zou2018potrojan}
Zou, M., Shi, Y., Wang, C., Li, F., Song, W., and Wang, Y.
\newblock Potrojan: powerful neural-level trojan designs in deep learning
  models.
\newblock \emph{arXiv preprint arXiv:1802.03043}, 2018.

\end{thebibliography}
%
\end{document}